\documentclass[10pt,twocolumn,a4paper]{article}
\usepackage[margin=20mm,columnsep=7mm]{geometry}
\usepackage{fontspec,xeCJK}
\setCJKmonofont[Path=fonts/]{NotoSansCJKsc-Regular.otf}
\usepackage{amsmath,amssymb,graphicx,booktabs,array,caption}
\usepackage{microtype,xurl,ragged2e,xcolor,flushend}
\definecolor{linkblue}{HTML}{5285B0}
\usepackage[colorlinks=true,allcolors=linkblue,pdfauthor={Frank Li},pdftitle={Validating Hybrid-State Cache Recovery for GLM-5.3-Flash with vLLM and LMCache}]{hyperref}
\newcommand{\paperCode}[1]{\mbox{\texttt{\detokenize{#1}}}}
\makeatletter
\renewcommand{\section}{\@startsection{section}{1}{\z@}{-2.8ex plus -1ex minus -.2ex}{1.3ex plus .2ex}{\normalfont\large\bfseries\raggedright}}
\renewcommand{\subsection}{\@startsection{subsection}{2}{\z@}{-2ex plus -.5ex minus -.2ex}{.8ex plus .2ex}{\normalfont\normalsize\bfseries\raggedright}}
\makeatother
\begin{document}

\raggedbottom
\twocolumn[
\begin{center}
{\LARGE\bfseries Validating Hybrid-State Cache Recovery for GLM-5.3-Flash with vLLM and LMCache\par}
\vspace{9pt}
{\large Frank Li\par}
\vspace{2pt}
{UNSW Sydney · \href{mailto:research@n1a.net}{research@n1a.net}\par}
\end{center}
\begin{minipage}{\textwidth}
\small
\noindent\textbf{Abstract.} External cache transfers can succeed while a hybrid language model resumes from an inconsistent state. We examine the full 45-layer GLM-5.3-Flash model, using the \paperCode{RedHatAI/}\allowbreak{}\paperCode{GLM-}\allowbreak{}\paperCode{5.}\allowbreak{}\paperCode{3-}\allowbreak{}\paperCode{Flash-}\allowbreak{}\paperCode{NVFP4} quantized checkpoint with vLLM and LMCache under four-way tensor parallelism. A complete-hit recovery mismatch restored state for the full prompt while the scheduler credited one fewer token. We aligned recovery through strict-prefix lookup and established a numerical comparison using shared computation corrections, matched checkpoint scheduling, and fixed per-rank kernel configurations. In a nine-length serial workload, agreement with the modified recomputation control improved from 34/36 to 36/36 generations, each containing 64 token IDs. A separate instrumented run passed recorded transfer-page, effective-tail, and delayed-save checks. Three additional synthetic templates passed 72 paired 256-token continuations across two fresh-container runs. A subsequent serial performance study preserved output equality across 120 requests; among the measured trials, CPU reload reduced time to first token by 46--64\% and total request time by 1.9--7.0\% relative to modified cold recomputation. The contribution is an experimentally validated integration repair applying an existing checkpoint-alignment principle. The evidence is confined to one model revision and controlled configuration; it does not establish general determinism, task-quality equivalence, concurrent-serving gains, or capacity beyond GPU memory.
\vspace{12pt}
\end{minipage}
]
\RaggedRight
\section{Introduction}

At a prompt length of 3,584 tokens, four transfer workers reported restoring a cached checkpoint for all 3,584 tokens. The scheduler, however, credited only 3,583 tokens and scheduled one additional input token. The subsequent continuation differed from the no-connector control at generated-token index 11. Transfer had occurred on every tensor-parallel rank; a positive cache-hit counter did not explain whether computation resumed from the correct logical position.

This distinction matters when a cache contains state updated in place as well as attention history. Reducing a token count cannot reconstruct an earlier recurrent state. Existing work already describes the need to align reusable attention prefixes and recurrent checkpoints; we apply that principle to a concrete integration failure rather than propose a new checkpointing algorithm~\cite{ref1}.

The failure was difficult to isolate because the output reference also required validation. Earlier paired runs could diverge despite identical recorded weights, and apparently successful byte checks had initially inspected ineffective all-zero tail locations. Neither source inspection nor one successful transfer was sufficient evidence. The eventual comparison therefore used a modified recomputation control with shared computation changes, a matched checkpoint schedule, and a fixed per-rank Flash Linear Attention (FLA) kernel profile. It did not compare a connector-only patch against an untouched production executable.

This case study addresses three questions: which recovery condition failed in the integration; how a useful numerical control was established; and what the repaired path demonstrably passed. Its contributions are an observed complete-hit state/position mismatch and its repair, a documented construction of the numerical comparison, and a bounded regression combining generated-token agreement with transfer and completion checks. The historical investigation contains 53 changing experimental rounds, not 53 independent repetitions of one treatment. The main evaluation uses the final, explicitly identified comparisons.

\section{System and comparison design}

\subsection{Execution and cached state}

The experiment runs one full GLM-5.3-Flash target model across four GPUs with tensor parallelism. GLM-5.3-Flash combines sparse and linear attention; the evaluated NVFP4 checkpoint is Red Hat's quantized derivative of the Z.ai model. Our workloads use text inputs only~\cite{ref8},~\cite{ref9}. Throughout this paper, ``GLM'' abbreviates this evaluated model and checkpoint, not the GLM model family as a whole. vLLM performs model execution and scheduling; LMCache provides the external cache path. The four ranks are parts of one inference engine, not four independent model replicas. The experiment does not implement prefill/decode disaggregation.

Figure 1 summarizes the execution and recovery responsibilities.

\begin{figure*}[!t]
\centering
\includegraphics[width=\textwidth]{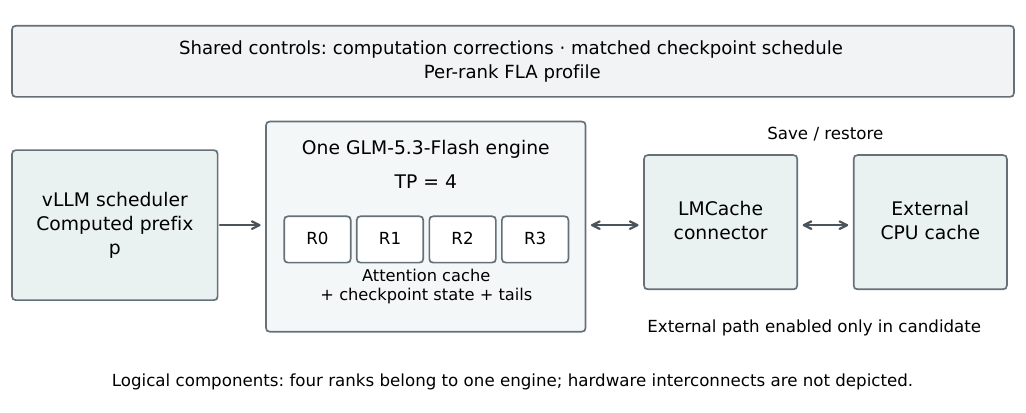}
\caption{Evaluated logical components. The four tensor-parallel ranks form one GLM-5.3-Flash engine. Both arms share modified computation controls; only the candidate enables the external CPU-cache path. Arrows describe logical interactions, not hardware topology.}
\label{fig:1}
\end{figure*}

The integration handles attention-related cache entries, checkpointed state, and model-specific tail entries. Logical engine groups and physical transfer groups are distinct: the audited registration contains 70 entries per rank across six transfer groups. Some transfer representations are byte views, while the observed tail entries are BF16. Thus neither the model checkpoint's NVFP4 name nor the FP8 KV configuration describes every state tensor's precision.

\begin{table*}[!t]
\centering\small
\caption{Evaluated configuration}
\label{tab:1}
\begin{tabular}{@{}>{\setlength{\RaggedRightParindent}{0pt}\RaggedRight\arraybackslash}p{\dimexpr 0.22000\textwidth-1.00000\tabcolsep\relax}>{\setlength{\RaggedRightParindent}{0pt}\RaggedRight\arraybackslash}p{\dimexpr 0.78000\textwidth-1.00000\tabcolsep\relax}@{}}
\toprule
\textbf{Setting} & \textbf{Evaluated configuration} \\
\midrule
Target & Full 45-layer \paperCode{RedHatAI/}\allowbreak{}\paperCode{GLM-}\allowbreak{}\paperCode{5.}\allowbreak{}\paperCode{3-}\allowbreak{}\paperCode{Flash-}\allowbreak{}\paperCode{NVFP4} \\
Model revision & \paperCode{36c184c6}\allowbreak{}\paperCode{cda000a4}\allowbreak{}\paperCode{81711306}\allowbreak{}\paperCode{df5adde4}\allowbreak{}\paperCode{2f63321a} \\
Runtime & vLLM \paperCode{0.}\allowbreak{}\paperCode{1.}\allowbreak{}\paperCode{dev20051+g487ecf187}; LMCache \paperCode{0.}\allowbreak{}\paperCode{5.}\allowbreak{}\paperCode{4}; PyTorch \paperCode{2.}\allowbreak{}\paperCode{13.}\allowbreak{}\paperCode{0+cu130} \\
Parallelism & TP=4, PP=1, DP=1; four GPUs on one host \\
Precision & Configured model dtype: BF16; \paperCode{compressed-}\allowbreak{}\paperCode{tensors} quantization; FP8 KV setting \\
Execution & Eager, synchronous scheduling; MTP configured with one speculative token \\
Limits & 4 GiB KV budget per rank; maximum model length 32,768; maximum sequences 4; token batch limit 8,192 \\
Checkpoint interval & C=1,792 tokens \\
Numerical control & Seven recorded FLA autotuner configurations per rank \\
\bottomrule
\end{tabular}
\end{table*}

MTP configuration does not imply that every step drafts a token: checkpoint-protection steps can disable drafting. The later content extension records four NVIDIA RTX PRO 6000 Blackwell Server Edition GPUs and their UUIDs. This current inventory does not retroactively establish the exact devices used in earlier rounds. Performance-specific runtime settings are reported in §5.5; the shared computation controls are described below.

\subsection{State position and scheduler position}

Let N denote prompt length, C the checkpoint interval, B the logical prefix represented by the restored checkpoint, and p the scheduler's computed-token count. Prefix lengths are counts; token indices are zero based. A state representing prefix {[}0,B) supports continuation from B. Crediting p=B therefore aligns the restored state with the start of subsequent computation.

For the fully cached prompts in the final workload, restricting lookup to the prompt without its last token selects B=C floor((N−1)/C). The scheduler then computes the nonempty suffix {[}B,N). This formula assumes the aligned checkpoint is available, as it was in these reload tests. In a general cache, lookup must select an available checkpoint within the query boundary; the formula is not a guarantee of a hit.

\begin{figure*}[!t]
\centering
\includegraphics[width=\textwidth]{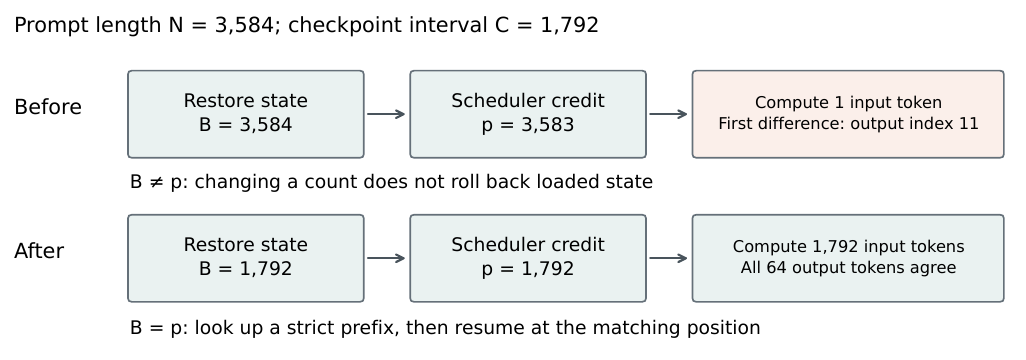}
\caption{Checkpoint alignment at N=3,584 and C=1,792. Before repair, the loaded state represents B=3,584 while the scheduler credits p=3,583. Strict-prefix lookup instead restores B=p=1,792 and recomputes the remaining suffix. This is a logical boundary example, not a timing trace.}
\label{fig:2}
\end{figure*}

\subsection{Treatments and observations}

\begin{table*}[!t]
\centering\small
\caption{Treatments and observation conditions}
\label{tab:2}
\begin{tabular}{@{}>{\setlength{\RaggedRightParindent}{0pt}\RaggedRight\arraybackslash}p{\dimexpr 0.22000\textwidth-1.50000\tabcolsep\relax}>{\setlength{\RaggedRightParindent}{0pt}\RaggedRight\arraybackslash}p{\dimexpr 0.35000\textwidth-1.50000\tabcolsep\relax}>{\setlength{\RaggedRightParindent}{0pt}\RaggedRight\arraybackslash}p{\dimexpr 0.16000\textwidth-1.50000\tabcolsep\relax}>{\setlength{\RaggedRightParindent}{0pt}\RaggedRight\arraybackslash}p{\dimexpr 0.27000\textwidth-1.50000\tabcolsep\relax}@{}}
\toprule
\textbf{Configuration} & \textbf{Computation and scheduling} & \textbf{Connector} & \textbf{Observation conditions} \\
\midrule
Original production configuration & Original production settings & Not the evaluated treatment & Not the output control in the final matrix \\
Modified recomputation control & Shared computation corrections, fixed per-rank FLA profile, matched checkpoint schedule & Disabled & CPU batch/configuration logging; no GPU tensor probes \\
Modified candidate & Corresponding common corrections and FLA profile; connector checkpoint path & Enabled; lookup boundary changed between the two matrices & CPU batch/configuration and transfer logging; no GPU tensor probes \\
Instrumented candidate & Repaired candidate configuration & Enabled & GPU byte checks, destination poisoning, tail checks, injected save delay \\
\bottomrule
\end{tabular}
\end{table*}

The cumulative functional patch changes eleven files across LMCache and vLLM. It includes layout and snapshot handling, save completion, checkpoint behavior, and stable selection/packing changes. The FLA installer and the no-connector scheduling fallback are separate controls. Consequently, successful evaluation is a result for their stated combination; it does not establish that the cumulative patch alone preserves production outputs. The final strict-prefix intervention changes two GLM lookup entry points within this existing configuration.

\section{Integration failures and repair}

\subsection{Observing effective state and waiting for saves}

Early tail checks reported equal all-zero pages. Later examination of the effective locations and strides invalidated that coverage claim: equality at an irrelevant address was not a test of the state consumed by subsequent computation. Correcting the addressing exposed nonzero tail data and transfer discrepancies that the earlier observation had missed. The earlier success interpretation was retained as a corrected failure of observation, not counted toward the final validation.

Save submission also had to be distinguished from completion. The integration was changed to wait on the actual \paperCode{DeviceMessagingFuture} before allowing dependent state reuse. The final regression inserted a 500 ms delay into saves and paired the actual waits with the delayed operations. This tests whether the recorded dependency is respected under the injected condition; it does not estimate a production failure probability.

These checks answer separate questions. Comparing source and destination pages checks the audited bytes. Observing nonzero effective tails checks that selected useful state entries are exercised. Waiting for the actual save checks a completion dependency. None alone establishes that the bytes describe the prefix from which the scheduler resumes.

\subsection{Complete-hit boundary mismatch}

The fixed-profile pre-fix matrix exposed failures at N=3,584 and N=5,376. For both lengths, raw transfer events from every rank reported a checkpoint covering the entire prompt, B=N. The scheduler's computed count and external-hit metric instead reported p=N−1, and the first reload batch scheduled one input token.

The archived connector explains the discrepancy. It retained the aligned lookup result in the request tracker used for snapshot retrieval. A later complete-prompt branch reduced the amount credited to the scheduler by one. Snapshot selection still used the retained full boundary. The observed condition was therefore B=N and p=N−1, rather than a snapshot of the shorter prefix. This mechanism is supported by the actual transfer and batch records, not inferred from the metric alone.

The repair excludes the final known token before lookup at both GLM entry points. Alignment then selects an earlier complete checkpoint for an exact-boundary prompt. Restricting the query before checkpoint selection also keeps selection and its associated lookup bookkeeping at the same boundary; subtracting a count after retrieving full state would leave the original problem intact.

At N=3,584, the repaired path loads B=1,792 and computes a 1,792-token suffix. At N=5,376, it loads B=3,584 and computes the same suffix length. Both previously failing continuations then match their controls. This repair increases recomputation relative to the faulty one-token schedule at exact boundaries. The incremental latency cost relative to that incorrect path has not been measured; §5.5 instead compares repaired reload with modified cold recomputation.

\section{Establishing a numerical control}

\subsection{What the diagnostic comparisons established}

Temperature zero and a fixed seed did not make the original cross-process comparison a reliable attribution test. Recorded batches differed: an earlier no-connector path processed a 3,583-token prefill, while the connector path split it into 1,792 and 1,791 tokens; decode scheduling differed as well. A matched checkpoint schedule removed this particular discrepancy, but output differences remained.

A subsequent audit found matching named parameter and buffer storages across arms. That narrowed the investigation without proving that all scratch allocations or execution choices were identical. An actual mHC invocation then showed that the attention-output input already differed between arms, while the other seven captured tensor inputs and scalar arguments agreed. This observation does not demonstrate a same-input mHC defect. Probes taken in separate runs also cannot be joined into one observed layer-by-layer causal trace.

The relevant prior distinction is between repeated execution at one shape and invariance across batching or prefix splits. Fixed seeds do not resolve changes in floating-point reduction geometry. He discusses this distinction and concrete kernel controls in an author technical report; those results motivate the local control but do not validate this GLM implementation~\cite{ref2}.

\subsection{Recorded FLA profile and its scope}

The experiment recorded actual FLA autotuner choices in a baseline run and pinned the seven observed configurations separately for each rank. The installer constrained each observed tuner to its recorded configuration and cleared its selection cache. Invocation of an unrecorded tuner was rejected. Later runs additionally checked the actual selected configuration after calls.

\begin{table*}[!t]
\centering\small
\caption{Numerical-control experiments}
\label{tab:3}
\begin{tabular}{@{}>{\setlength{\RaggedRightParindent}{0pt}\RaggedRight\arraybackslash}p{\dimexpr 0.23000\textwidth-1.33333\tabcolsep\relax}>{\setlength{\RaggedRightParindent}{0pt}\RaggedRight\arraybackslash}p{\dimexpr 0.40000\textwidth-1.33333\tabcolsep\relax}>{\setlength{\RaggedRightParindent}{0pt}\RaggedRight\arraybackslash}p{\dimexpr 0.37000\textwidth-1.33333\tabcolsep\relax}@{}}
\toprule
\textbf{Experiment} & \textbf{Observation} & \textbf{Interpretation} \\
\midrule
Profile capture & Recorded per-rank configurations and actual call inputs & A concrete reference for the next intervention \\
\addlinespace[1.5pt]
Fixed-profile first case & All four 64-token generations agree between arms and with the selected reference & Supports the controlled paired comparison \\
\addlinespace[1.5pt]
Fresh-container repeat & New engines reproduce those first-case comparisons & Independent repetition of that bounded test \\
\addlinespace[1.5pt]
Full boundary matrices & Actual configurations verified in both arms on four ranks & Numerical control retained during the boundary intervention \\
\bottomrule
\end{tabular}
\end{table*}

The profile was a joint intervention on seven configurations per rank. No single-kernel ablation establishes that one tile size, warp count, or kernel alone caused the earlier divergence. The profile is neither a general determinism algorithm nor an established performance optimum.

An older reference from a different diagnostic round remained unequal under the fixed profile. Its failed assertion and nonzero process exit were preserved. The selected profile reference was identified before the intervention; agreement with it does not turn the older comparison into a pass. This distinction prevents a changing numerical reference from silently changing the reported acceptance criterion.

\section{Evaluation}

\subsection{Workload and metrics}

The same archived plan is used for the pre-fix, repaired, and instrumented matrices. It contains nine prompt lengths around selected boundaries: 3,583, 3,584, 3,585, 3,587, 3,588, 3,589, 5,375, 5,376, and 5,377. Each case performs five serial operations: generate the target from a cold state, generate two interposer requests, reset the GPU cache while retaining the external cache, and regenerate the target. Each paired arm therefore executes 45 operations, including 36 generations and nine resets.

The inputs comprise 27 distinct token sequences generated from one Chinese password-retrieval and explanation template. Target and interposer variants differ in a document tag near the start; they are not independent application domains. There is no explicit request cache salt, and recorded transfer events have an empty salt. The output criterion compares all 64 generated token IDs per request, with temperature zero, seed 42, and EOS ignored. No recorded output contains the known EOS token IDs. Although the prompt requests a longer explanation, the 64-token cap is a finite continuation test, not evidence of satisfying that instruction or preserving task quality.

We distinguish paired output equality, within-engine cold/reload equality, actual four-rank transfers, and internal transfer observations. There are 36 paired comparisons in each two-arm matrix, not 36 distinct tasks. Preemption counters remain zero. The serial reset/interposer procedure exercises the cache path; it does not demonstrate behavior under sustained memory pressure or concurrent preemption.

Evidence checks separately recompute results from archived records and cross-check recorded observations; they do not provide an independently implemented reference for every low-level checker. Hash agreement establishes byte identity, not correct state addressing. The retrospective page audit verifies recorded comparisons rather than re-reading historical device buffers.

\subsection{Boundary results without GPU tensor probes}

Table 4 retains all nine cases. ``First difference'' is the zero-based generated-token index in the pre-fix reload comparison; a dash means the complete 64-token continuation agrees. B comes from transferred checkpoint metadata and p from the first reload batch. Each repaired row contains four paired generations: target cold, two interposers, and target reload.

\begin{table*}[!t]
\centering\small
\caption{Complete boundary matrix before and after repair}
\label{tab:4}
\begin{tabular}{@{}>{\setlength{\RaggedRightParindent}{0pt}\RaggedRight\arraybackslash}p{\dimexpr 0.09000\textwidth-1.71429\tabcolsep\relax}>{\setlength{\RaggedRightParindent}{0pt}\RaggedRight\arraybackslash}p{\dimexpr 0.12000\textwidth-1.71429\tabcolsep\relax}>{\setlength{\RaggedRightParindent}{0pt}\RaggedRight\arraybackslash}p{\dimexpr 0.12000\textwidth-1.71429\tabcolsep\relax}>{\setlength{\RaggedRightParindent}{0pt}\RaggedRight\arraybackslash}p{\dimexpr 0.19000\textwidth-1.71429\tabcolsep\relax}>{\setlength{\RaggedRightParindent}{0pt}\RaggedRight\arraybackslash}p{\dimexpr 0.16000\textwidth-1.71429\tabcolsep\relax}>{\setlength{\RaggedRightParindent}{0pt}\RaggedRight\arraybackslash}p{\dimexpr 0.15000\textwidth-1.71429\tabcolsep\relax}>{\setlength{\RaggedRightParindent}{0pt}\RaggedRight\arraybackslash}p{\dimexpr 0.17000\textwidth-1.71429\tabcolsep\relax}@{}}
\toprule
\textbf{N} & \textbf{B before} & \textbf{p before} & \textbf{First difference before} & \textbf{B = p after} & \textbf{Suffix after} & \textbf{Equal generations after} \\
\midrule
3583 & 1792 & 1792 & --- & 1792 & 1791 & 4/4 \\
3584 & 3584 & 3583 & 11 & 1792 & 1792 & 4/4 \\
3585 & 3584 & 3584 & --- & 3584 & 1 & 4/4 \\
3587 & 3584 & 3584 & --- & 3584 & 3 & 4/4 \\
3588 & 3584 & 3584 & --- & 3584 & 4 & 4/4 \\
3589 & 3584 & 3584 & --- & 3584 & 5 & 4/4 \\
5375 & 3584 & 3584 & --- & 3584 & 1791 & 4/4 \\
5376 & 5376 & 5375 & 0 & 3584 & 1792 & 4/4 \\
5377 & 5376 & 5376 & --- & 5376 & 1 & 4/4 \\
\bottomrule
\end{tabular}
\end{table*}

The pre-fix matrix has 34/36 equal generation pairs, with the two differences confined to the exact-boundary reloads. The repaired matrix has 36/36 equal pairs, and all nine target cold/reload comparisons agree. All 36 control generations also agree between the pre-fix and repaired runs. Across the repaired matrix's nine reloads and four ranks, the first batch has computed=B and scheduled=N−B. Every reload records positive H2D transfer on all four ranks.

At N=3,584, transferred bytes per rank change from 69,625,856 to 54,809,600; at N=5,376, they change from 84,442,112 to 69,625,856. These metadata corroborate the change in selected checkpoints. These before/after byte counts alone do not establish a latency benefit. Section 5.5 separately measures the repaired path against modified cold recomputation; it is not a performance comparison against the incorrect pre-fix path.

An earlier attempt to execute the matrix inadvertently ran only the five-operation first case because of a harness import entry point. The independent coverage verifier rejected it. It is excluded from the nine-case results. Likewise, the changing diagnostic rounds are not pooled into a statistical reliability estimate.

\subsection{Instrumented transfer and save regression}

A separate repaired candidate executes the complete plan with byte observers, destination poisoning before H2D, and injected save delay. Its 36 generations match the archived repaired-run control, and all nine cold/reload pairs agree. This is an instrumented candidate compared with a previously recorded control, not a new two-arm run.

\begin{table*}[!t]
\centering\small
\caption{Instrumented transfer and save regression}
\label{tab:5}
\begin{tabular}{@{}>{\setlength{\RaggedRightParindent}{0pt}\RaggedRight\arraybackslash}p{\dimexpr 0.30000\textwidth-1.33333\tabcolsep\relax}>{\setlength{\RaggedRightParindent}{0pt}\RaggedRight\arraybackslash}p{\dimexpr 0.34000\textwidth-1.33333\tabcolsep\relax}>{\setlength{\RaggedRightParindent}{0pt}\RaggedRight\arraybackslash}p{\dimexpr 0.36000\textwidth-1.33333\tabcolsep\relax}@{}}
\toprule
\textbf{Check} & \textbf{Observed coverage} & \textbf{Result} \\
\midrule
Candidate/control generations & 36 continuations × 64 tokens & All agree \\
\addlinespace[1.5pt]
D2H page-comparison rows & 17,640 & Zero reported byte mismatches \\
\addlinespace[1.5pt]
H2D page-comparison rows & 3,288 & Zero reported byte mismatches; destinations poisoned \\
\addlinespace[1.5pt]
Combined page-comparison rows & 20,928 & All 70 registered entries covered per rank and direction \\
\addlinespace[1.5pt]
Nonzero tail H2D observations & 4 ranks × 9 reloads × 12 entries = 432 & All observed entries nonzero \\
\addlinespace[1.5pt]
Delayed-save wait pairs & 67 per rank; 268 total & Actual waits cover the injected 500 ms delay \\
\addlinespace[1.5pt]
H2D volume & 2,447,265,792 bytes across four ranks & Positive transfers for every reload \\
\bottomrule
\end{tabular}
\end{table*}

Page rows are repeated internal checks, not 20,928 independent round trips or randomized trials. Nonzero tails demonstrate coverage of the selected entries, not equality of every internal model state. The artificial delay and heavy observation deliberately change execution conditions, so this run is not a serving-performance measurement. Its agreement complements the output matrix without GPU tensor probes; it does not replace that matrix.

\subsection{Additional prompt templates and longer continuations}

A subsequent preregistered extension tests three additional synthetic templates: English ledger arithmetic, Chinese ordered-rule application, and English Python-code reasoning. Each template uses prompt lengths 3,583, 3,584, and 5,376, with the same cold/interposer/reset/reload procedure and 256 generated tokens per new request. The historical first case remains as a 64-token bridge to the previously selected reference. The combined plan contains 30 distinct input sequences and 50 operations per arm: four bridge generations and 36 extended generations, totaling 9,472 output tokens per arm.

Two separate disposable containers each run a fresh no-connector engine and a fresh candidate engine under the existing fixed configuration. The exported token plans are identical. Both pairs pass all 40 continuation comparisons, and the two runs also agree on every baseline and candidate continuation. For all ten reloads per run, four-rank transfer metadata and first-batch records agree on B=C floor((N−1)/C), computed=B, scheduled=N−B, and zero draft tokens in the protected first batch. Actual FLA profiles match the recorded policy; no known EOS token or request preemption is observed. Each run exports 49 files whose SHA hashes are independently verified.

\begin{table*}[!t]
\centering\small
\caption{Content and continuation extension across fresh engines}
\label{tab:6}
\begin{tabular}{@{}>{\setlength{\RaggedRightParindent}{0pt}\RaggedRight\arraybackslash}p{\dimexpr 0.60000\textwidth-1.33333\tabcolsep\relax}>{\setlength{\RaggedRightParindent}{0pt}\RaggedRight\arraybackslash}p{\dimexpr 0.20000\textwidth-1.33333\tabcolsep\relax}>{\setlength{\RaggedRightParindent}{0pt}\RaggedRight\arraybackslash}p{\dimexpr 0.20000\textwidth-1.33333\tabcolsep\relax}@{}}
\toprule
\textbf{Measure} & \textbf{First pair} & \textbf{Fresh repeat} \\
\midrule
Operations per arm & 50 & 50 \\
Historical 64-token bridge comparisons & 4/4 & 4/4 \\
New 256-token comparisons & 36/36 & 36/36 \\
Output tokens per arm & 9,472 & 9,472 \\
Four-rank reload cases with aligned checkpoint and batch & 10/10 & 10/10 \\
Known EOS occurrences, both arms & 0 & 0 \\
Request preemptions, both arms & 0 & 0 \\
\bottomrule
\end{tabular}
\end{table*}

This adds 72 successful 256-token generation pairs and eight repeated bridge pairs across two container-level repetitions. These are repeated synthetic inputs, not 80 independent tasks. A setup attempt was aborted before workload execution after detecting an incorrect served-model identifier in the new harness; it has no model-correctness verdict and is excluded. The prompt contents, continuation lengths, output reference, and acceptance criteria were not changed in response to model results. This extension does not repeat the poisoned byte audit, establish task-answer quality, or measure performance. The complete plan, failed setup record, raw exports, and independent checks are retained in the internal evidence archive.

\subsection{Paired serial performance}

A separately preregistered study measures serial streaming completions on the disposable container's loopback interface. Two fresh containers run the same modified controls in opposite orders: OFF→ON, then ON→OFF. Each engine runs one excluded warmup and three measured trials per length, N\ensuremath{\in}\{3583,3584,5376\}, with distinct early document identities and 128-token continuations. OFF attempts cold and immediate-repeat requests; ON additionally attempts CPU reload after a GPU-only reset. All 120 requests (90 measured) pass complete output-token checks: each input has the same output across its five request conditions and both containers. The four engines are two paired repetitions, not 120 independent experiments. The workload uses one English ledger task (opening balance 137, receipt 58, dispatch 29) with repeated background text. Its 12 tokenized inputs vary length and early document identity; nine inputs enter the measured sample. They are input variants of one synthetic task, not 12 distinct tasks. Requests use temperature 0, seed 42, and ignore\_eos=True; task-answer quality is not evaluated.

TTFT measures request start to receipt of the first nonempty token-ID event; total latency ends after the stream's DONE marker. Decode rate counts tokens after the first event over the first-to-last nonempty-event interval, allowing multiple tokens per event. CPU metadata observers remain enabled; GPU tensor probes, poisoned destinations, artificial save delays, initialization and inter-request pauses are excluded. The GPUs are dedicated to the experiment while other production workloads continue on the shared host. Trials run in fixed ascending order of trial index and prompt length; only engine order is reversed. This balances engine order across two runs but does not randomize request history or eliminate shared-host interference.

The performance runs use four NVIDIA RTX PRO 6000 Blackwell Server Edition GPUs. Both arms use the Marlin MoE backend with custom all-reduce disabled. NCCL\_P2P\_DISABLE is set to 1 and OMP\_NUM\_THREADS to 2; expandable\_segments is False. FlashInfer autotuning is disabled while the separately fixed FLA profile remains in force. The candidate's local LMCache server uses LRU, separate object groups, l1-size-gb=16, and disabled lazy L1 allocation. These settings bound the performance result; no inter-host cache transfer or optimized communication comparison is measured.

\begin{figure*}[!t]
\centering
\includegraphics[width=\textwidth]{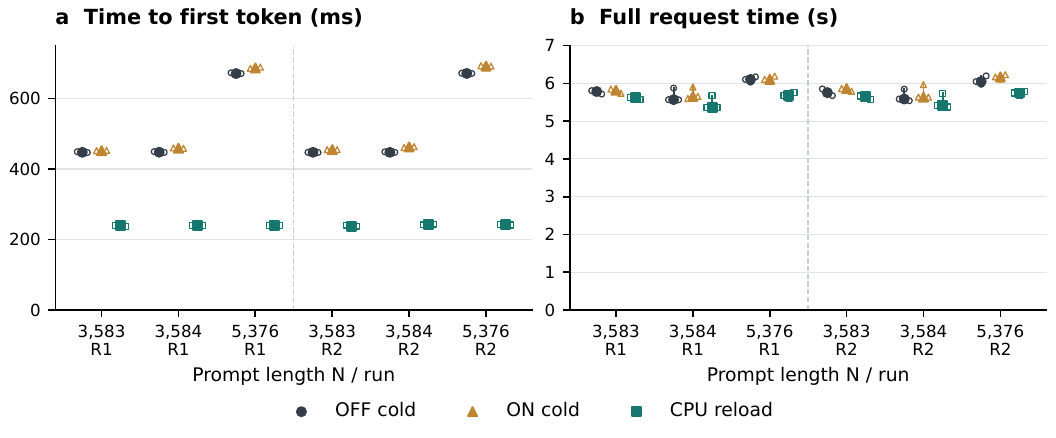}
\caption{Serial latency measurements. Hollow points show three measured requests per run, length, and condition; filled markers show medians, and thin bars show observed min--max ranges, not confidence intervals. R1 runs OFF then ON; R2 reverses engine order. Both panels use the same 54 requests and zero-based axes. OFF is the modified no-connector control. The cold conditions have similar TTFT; total time includes the full 128-token generation.}
\label{fig:3}
\end{figure*}

\begin{table*}[!t]
\centering\small
\caption{Serial latency under matched controls}
\label{tab:7}
\begin{tabular}{@{}>{\setlength{\RaggedRightParindent}{0pt}\RaggedRight\arraybackslash}p{\dimexpr 0.07000\textwidth-1.71429\tabcolsep\relax}>{\setlength{\RaggedRightParindent}{0pt}\RaggedRight\arraybackslash}p{\dimexpr 0.09000\textwidth-1.71429\tabcolsep\relax}>{\setlength{\RaggedRightParindent}{0pt}\RaggedRight\arraybackslash}p{\dimexpr 0.18000\textwidth-1.71429\tabcolsep\relax}>{\setlength{\RaggedRightParindent}{0pt}\RaggedRight\arraybackslash}p{\dimexpr 0.18000\textwidth-1.71429\tabcolsep\relax}>{\setlength{\RaggedRightParindent}{0pt}\RaggedRight\arraybackslash}p{\dimexpr 0.18000\textwidth-1.71429\tabcolsep\relax}>{\setlength{\RaggedRightParindent}{0pt}\RaggedRight\arraybackslash}p{\dimexpr 0.15000\textwidth-1.71429\tabcolsep\relax}>{\setlength{\RaggedRightParindent}{0pt}\RaggedRight\arraybackslash}p{\dimexpr 0.15000\textwidth-1.71429\tabcolsep\relax}@{}}
\toprule
\textbf{Run} & \textbf{N} & \textbf{OFF cold TTFT (ms)} & \textbf{ON cold TTFT (ms)} & \textbf{ON reload TTFT (ms)} & \textbf{TTFT ratio} & \textbf{Total ratio} \\
\midrule
1 & 3,583 & 447.3 & 451.9 & 239.4 & 1.87 & 1.030 \\
1 & 3,584 & 447.6 & 458.8 & 240.4 & 1.86 & 1.037 \\
1 & 5,376 & 670.9 & 686.7 & 239.7 & 2.80 & 1.076 \\
2 & 3,583 & 447.3 & 455.0 & 237.9 & 1.88 & 1.019 \\
2 & 3,584 & 447.2 & 462.6 & 241.4 & 1.85 & 1.031 \\
2 & 5,376 & 671.2 & 691.8 & 241.2 & 2.78 & 1.053 \\
\bottomrule
\end{tabular}
\end{table*}

Each latency is the median of three measured trials. Ratios are medians of per-input OFF-cold/ON-reload ratios, not ratios of the displayed medians. Across the six run/length groups, TTFT ratios range from 1.85--2.80 and total-latency ratios from 1.019--1.076. ON-cold/OFF-cold TTFT ratios range from 1.01--1.04; this is an end-to-end enabled/disabled comparison under the shared modified controls. It does not isolate transfer latency, observer cost, or the cost of all changes relative to production. Raw per-request values, ranges and event-level decode rates are retained in the evidence tables.

Across these same six groups, OFF-cold total-latency medians span 5.570--6.104 s and ON-reload medians span 5.369--5.745 s. Post-first-event decode-rate medians span 23.38--24.80 and 23.08--24.76 tokens/s, respectively. These ranges describe the observed time scale across inputs and runs; they are not uncertainty intervals.

Attempted path names do not determine actual paths. Among 18 measured OFF immediate repeats, 12 recompute and 6 reuse a local prefix; among 18 ON immediate repeats, 18 reload from CPU and 0 use local-only reuse. The OFF local reuse occurs only at N=5376 and credits 1792 tokens; its N=3583 and N=3584 repeats recompute. At N=5376 the ON CPU path instead restores 3584 tokens, so those repeat conditions do not perform equal suffix work. Consequently, a complete five-path local-hit comparison is unavailable. This is an observed limitation of the evaluated configuration, not evidence that hybrid models cannot support local reuse. Each of the 18 measured explicit CPU reloads has positive four-rank H2D and an independently verified first-batch prefix B=p: 1792, 1792 and 3584 tokens for the three lengths, leaving 1791, 1792 and 1792 prompt tokens to recompute. Classification retains this actual suffix work. We report no confidence interval from two engine pairs, sustained-concurrency result, or capacity gain beyond GPU memory.

\section{Related work and limitations}

Hybrid-state reuse is an established problem. Marconi explains why recurrent states updated in place require checkpoints aligned with reusable attention prefixes and studies cache admission and eviction. The Sparse Prefix Caching preprint studies sparse checkpoint materialization and suffix recomputation for hybrid and recurrent serving. Our strict-prefix lookup repair applies this existing requirement to a connector/scheduler disagreement. We introduce neither a new eviction policy nor the principle of replaying a suffix from a checkpoint~\cite{ref1},~\cite{ref3}.

An upstream vLLM checkpoint RFC separately discusses prefix identity, non-KV logical state, and immutable payload versus logical equivalence. It is a design proposal, currently closed as not planned, rather than an implemented standard. Its prior discussion limits a novelty claim about a general recovery interface; the contribution here is evidence from a running integration~\cite{ref4}.

Numerical-control work is also directly relevant. He distinguishes repeatability from batch invariance. The LLM-42 preprint uses verified speculation and replaces speculative KV state with verified state even when generated tokens agree. These works reinforce why finite token equality should not be promoted to complete state or future-output identity. Our experiment fixes one observed profile and tests finite continuations; it does not implement a general online verification system~\cite{ref2},~\cite{ref5}.

LMCache supplies the cache infrastructure used in this study. Its broader cache management and transfer mechanisms are existing platform contributions. The local cumulative patch spans representation, completion, scheduling, and common computation changes; eleven changed files do not constitute eleven independently novel mechanisms. Only the final lookup intervention is isolated by the otherwise controlled before/after matrix. The preceding integration changes were not subjected to a full factorial ablation~\cite{ref6}.

External validity is limited by one model revision, one four-rank setup, synthetic prompt templates, finite continuations, serial requests, and a fixed profile selected from an observed baseline. The original matrix uses one template and 64-token outputs; the extension adds three templates and 256-token outputs, with only two fresh-container repetitions. We have not measured all logits, arbitrary future continuation, cross-TP invariance, concurrent preemption, other models, or default-production equivalence. The complete modified control is essential to interpreting the result.

The serial study measures a CPU-reload latency benefit against the shared modified cold control. Costs of fixed configuration, ordering changes and completion barriers relative to the original production configuration remain unmeasured. Immediate repeats do not consistently provide local-only reuse, limiting a complete five-path comparison. Two engine pairs on one shared host do not establish concurrent SLOs, broad reliability, or capacity beyond GPU memory. No public artifact release or production deployment is claimed.

\section{Conclusion}

The evaluated GLM integration restored a complete checkpoint while crediting the scheduler with a shorter prefix. Actual transfer metadata and batch records exposed that mismatch; restricting lookup to a strict prefix repaired both failing exact-boundary cases under a controlled numerical configuration. The repaired nine-case matrix passes all 36 finite continuation comparisons, and a separate instrumented run passes the specified byte, nonzero-tail, and delayed-save checks. The subsequent content extension also passes two fresh-container comparisons with 256-token continuations. The engineering result is a validated recovery path within these conditions, with a measured serial CPU-reload latency benefit under matched controls; representative concurrency, original-production overhead and overflow capacity remain open.

\par\addvspace{12pt plus 2pt minus 2pt}\noindent\begin{minipage}{\columnwidth}\setlength{\RaggedRightParindent}{0pt}\RaggedRight\setlength{\parskip}{3pt}

\section*{Acknowledgements}

The author acknowledges Research Technology Services, UNSW Sydney, for providing GPU resources on the Katana computational cluster~\cite{ref7} for model quantization.

\textbf{AI assistance.} AI tools assisted with the research and preparation of this manuscript. The author takes responsibility for the methods, results, and final text.

\par\end{minipage}

\renewcommand{\refname}{References}

\end{document}